\documentclass{elsart4-1}

\usepackage[dvips]{graphicx}
\usepackage{amssymb}

\usepackage[english,francais]{babel}

\newtheorem{e-proposition}[theorem]{Proposition}

\newtheorem{e-definition}[theorem]{Definition\rm}

\renewcommand{\theequation}{\arabic{equation}}
\def\og{\leavevmode\raise.3ex\hbox{$\scriptscriptstyle\langle\!\langle$~}}
\def\fg{\leavevmode\raise.3ex\hbox{~$\!\scriptscriptstyle\,\rangle\!\rangle$}}

\begin{document}

\begin{frontmatter}


% \author{Name\thanksref{label2}}
\selectlanguage{english}
\title{Scaling laws of turbulent dynamos}

\vspace{-2.6cm}

\selectlanguage{francais}
\title{Comportements asymptotiques des dynamos turbulentes}


\selectlanguage{english}
\author{St\'ephan Fauve, Fran\c cois P\'etr\'elis}
%\ead{fauve@lps.ens.fr}
%\author[authorlabel2]{Fran\c cois P\'etr\'elis}
%\ead{petrelis@lps.ens.fr}

\address{LPS, CNRS UMR 8550, ENS 24 rue Lhomond 75005 Paris France}
%\address[authorlabel2]{LPS, CNRS UMR 8550, ENS 24 rue Lhomond 75005 Paris France}

\begin{abstract}
% Text of abstract in English
{\it 
We consider magnetic fields generated by homogeneous isotropic and parity invariant turbulent flows. We show that simple scaling laws for dynamo threshold, magnetic energy and Ohmic dissipation can be obtained depending on the value of the  magnetic Prandtl number.}

\vskip 0.5\baselineskip

%\selectlanguage{francais}
%% Text of abstract in French
%\noindent{\bf R\'esum\'e}
%\vskip 0.5\baselineskip
%\noindent
 %{\it Nous \'etudions le comportement asymptotique des dynamos engendr\'ees par un \'ecoulement turbulent homog\`ene isotrope et invariant par sym\'etrie plane. Nous montrons que diff\'erents comportements asymptotiques simples pour le seuil, l'\'energie magn\'etique et la dissipation Joule peuvent \^etre obtenus en fonction de la valeur du nombre de Prandtl magn\'etique.}

%Now keywords/mots-clÈs
%\keywords dynamo; turbulence; magnetic field
%\vskip 0.5\baselineskip
\noindent{\small{\it keywords}~: dynamo~; turbulence~; magnetic field}
%\noindent{\small{\it Mots-cl\'es}~: dynamo~; turbulence~; champ magn\'etique}
\end{abstract}

\end{frontmatter}

% now the Version franÁaise abrÈgÈe, if it exists
\selectlanguage{francais}
\section*{Version fran\c{c}aise abr\'eg\'ee}
% Text of your Version franÁaise abrÈgÈe here
Il est \`a pr\'esent admis que les champs magn\'etiques des \'etoiles voire m\^eme des galaxies sont engendr\'es par l'\'ecoulement de fluides conducteurs de l'\'electricit\'e \cite{Zeldovitch,Widrow,Brandenburg}. Ceux-ci impliquent des nombres de Reynolds cin\'etique, $Re$, et magn\'etique, $R_m$, tr\`es \'elev\'es ($Re = VL/\nu$, $R_m = \mu_0 \sigma VL$, o\`u $V$ est l'\'ecart-type des fluctuations de vitesse, $L$, l'\'echelle int\'egrale de l'\'ecoulement, $\nu$, la viscosit\'e cin\'ematique du fluide, $\sigma$, sa conductivit\'e \'electrique et $\mu_0$, la perm\'eabilit\'e magn\'etique). Aucune exp\'erience de laboratoire ou simulation num\'erique directe des \'equations de la magn\'etohydrodynamique, ne permet l' \'etude du probl\`eme dans des r\'egimes de param\`etres, $Re$ et $R_m$,  d'int\'er\^et astrophysique. Il est donc utile de consid\'erer des hypoth\`eses plausibles afin de pousser plus loin l'analyse dimensionnelle qui, \`a partir des param\`etres $V$, $L$, $\nu$,  $\sigma$, $\mu_0$ et de la densit\'e du fluide $\rho$,  pr\'edit pour le seuil de l'effet dynamo et la densit\'e moyenne d'\'energie magn\'etique, $B^2 / 2\mu_0$, satur\'ee non lin\'eairement au-del\`a du seuil,
\begin{equation}
R_m^{c} = f(Re),
\label{thresholda}
\end{equation}
\begin{equation}
\frac{B^2}{\mu_0} = \rho V^2 \,g(R_m, Re).
\label{saturationa}
\end{equation}

Dans le cas d'un \'ecoulement turbulent homog\`ene isotrope, donc de vitesse moyenne nulle, et invariant par sym\'etrie plane, donc sans h\'elicit\'e, les r\'esultats des simulations num\'eriques les plus performantes r\'ealis\'ees \`a ce jour montrent que $R_m^{c}$ augmente continuellement en fonction de $Re$ \cite{Schekochihin}. Schekochihin et al. proposent que deux sc\'enarios extr\^emes, sch\'ematis\'ees dans la figure 1,  seront susceptibles d'\^etre observ\'es lorsque les ordinateurs auront acquis la puissance requise pour effectuer des calculs \`a $Re$ plus \'elev\'e: (i) une saturation $R_m^{c} \rightarrow \rm{constante}$, ou alors (ii) une croissance de la forme $R_m^{c} \propto Re$. D'autres simulations num\'eriques directes, r\'ealis\'ees avec des \'ecoulements turbulents poss\'edant un champ de vitesse moyen de g\'eom\'etrie fix\'ee ${\overline{\bf v} (\bf { r})} \neq 0$, semblent suivre le sc\'enario (i) \cite{PontyLaval}. 

Lorsque le nombre de Prandtl magn\'etique, $P_m = R_m / Re = \mu_0 \sigma \nu$, est faible, $P_m \ll 1$,  l'\'echelle de dissipation Joule du champ magn\'etique, $l_{\sigma} = L R_m^{-3/4}$, est grande par rapport \`a l'\'echelle de Kolmogorov $l_K = L Re^{-3/4}$. Le champ magn\'etique se d\'eveloppe donc \`a une \'echelle suffisamment grande pour ne pas \^etre affect\'e par la viscosit\'e cin\'ematique. Cette hypoth\`ese, couramment effectu\'ee en turbulence, permet de conclure en faveur du sc\'enario (i). En effet, si $\nu$ n'est pas pris en compte, l'analyse dimensionnelle impose $R_m^{c} \rightarrow \rm{constante}$. Il n'est donc pas surprenant que les mod\'elisations num\'eriques des grandes \'echelles, qui ne r\'esolvent pas les \'echelles dissipatives, donnent ce r\'esultat. 
Le sc\'enario (i) sera donc toujours observ\'e \`a $P_m$ suffisamment faible sous r\'eserve bien s\^ur que l'on ait dynamo. 

Il est cependant utile d'analyser le sc\'enario (ii) d'autant plus que, comme nous pouvons le remarquer, il correspond \`a la pr\'ediction faite par Batchelor en 1950 \cite{Batchelor}. En se basant sur une analogie entre l'\'equation de l'induction et celle de la vorticit\'e, Batchelor avait estim\'e que le seuil d'une dynamo engendr\'ee par un \'ecoulement turbulent devait correspondre \`a $P_m$ d'ordre unit\'e, soit $R_m^{c} \propto Re$. M\^eme si nous savons aujourd'hui que l'analyse de Batchelor est discutable, il est int\'eressant de d\'eterminer sous quelle hypoth\`ese minimale sa pr\'ediction est correcte. Supposons donc que nous nous limitions aux modes instables de champ magn\'etique, suffisamment localis\'es au sein de l'\'ecoulement afin de ne pas \^etre affect\'es par les conditions aux limites. Il est alors possible de ne pas prendre en compte l'\'echelle spatiale $L$, et l'analyse dimensionnelle impose pour le seuil, $P_m = \rm{constante}$, soit le sc\'enario (ii) $R_m^{c} \propto Re$. 
 
Les sc\'enarios consid\'er\'es ci-dessus conduisent aussi \`a des pr\'edictions diff\'erentes pour la densit\'e d'\'energie magn\'e\-tique engendr\'ee par effet dynamo. Le scenario (i) qui consiste \`a ne pas prendre en compte $\nu$ revient \`a n\'egliger la d\'ependance en $Re$ de $g(R_m, Re)$ dans (\ref{saturationa}). Au voisinage du seuil, $V$ est d\'etermin\'e par $R_m^{c} \sim \mu_0 \sigma V L$ et $g(Rm) \propto R_m - R_m^{c}$ dans le cas d'une bifurcation supercritique. Il en r\'esulte \cite{Petrelis01}
\begin{equation}
B^2 \propto \frac{ \rho} {\mu_0 (\sigma L)^2} (R_m-R_m^{c}).
\label{saturationseuila}
\end{equation}
Loin du seuil pour $P_m \ll 1$, $Re \gg R_m \gg R_m^{c}$, on peut supposer que $B$ ne d\'epend plus de $\sigma$ \`a condition que le champ magn\'etique se d\'eveloppe \`a une \'echelle plus grande que $l_{\sigma}$. Il en r\'esulte alors l'\'equipartition entre \'energie magn\'etique et cin\'etique, $B^2/\mu_0 \propto \rho V^2$, tel que suppos\'e initialement par Biermann et Schl\"uter \cite{Biermann}. 

Un r\'esultat compl\`etement diff\'erent est obtenu dans le sc\'enario (ii). Il convient de consid\'erer les param\`etres du probl\`eme sous la forme \'equivalente, $B$, $\epsilon = V^3/L$, $L$, $\nu$,  $\sigma$, $\mu_0$ et $\rho$. En effet, le champ magn\'etique \`a petite \'echelle est aliment\'e par la puissance par unit\'e de masse $\epsilon$ qui cascade depuis l'\'echelle int\'egrale, et il est donc important de conserver ce param\`etre m\^eme si l'on ne prend pas en compte explicitement $L$. L'analyse dimensionnelle conduit alors \`a 
\begin{equation}
\frac{B^2}{\mu_0} = \rho \sqrt{\nu \epsilon} \, h(P_m) = \frac{\rho V^2}{\sqrt{Re}} \, h(P_m),  
\label{saturationbatchelora}
\end{equation}
qui, pour $P_m \sim 1$, n'est autre que le r\'esultat obtenu par Batchelor en supposant que la saturation correspond \`a  l'\'equipartition entre l'\'energie magn\'etique et l'\'energie cin\'etique \`a l'\'echelle de Kolmogorov.

Revenons au cas $P_m \ll 1$ qui correspond aux \'ecoulements de m\'etaux liquides et plasmas \`a l'origine du champ magn\'etique des plan\`etes et des \'etoiles ($P_m < 10^{-5}$).
Dans ce cas, le champ magn\'etique se d\'eveloppe \`a des \'echelles a priori comprises entre $L$ et $l_{\sigma}$ avec $l_{\sigma} \gg l_K$ et il en r\'esulte que $R_m^{c}$ ne d\'epend pas de variations de $P_m$ (ou de $Re$) et que $B^2/\mu_0 = \rho V^2 g(R_m)$ (sc\'enario (i)). Int\'eressons nous \`a la puissance dissip\'ee par effet Joule par une telle dynamo. Il faut \`a cet effet d\'eterminer \`a quelles \'echelles se d\'eveloppe le champ magn\'etique. Utilisons pour cela un argument \`a la Kolmogorov en supposant que dans la zone inertielle, c'est \`a dire pour les nombres d'onde $k$ tels que $k l_{\sigma} \ll 1 \ll kL$, la puissance spectrale $\vert \hat B(k) \vert^2$ est ind\'ependante de $L$, $\sigma$ and $\nu$. Il en r\'esulte
\begin{equation}
\vert \hat B \vert^2 \propto \mu_0 \rho \,  \epsilon^{\frac{2}{3}} \, k^{-\frac{5}{3}}.
\label{kolmogorova}
\end{equation}
Ceci n'est pas la seule possibilit\'e parmi les nombreuses pr\'edictions relatives au spectre de la turbulence magn\'eto\-hydrodynamique, mais dans le cas pr\'esent, c'est probablement la plus simple. L'int\'egration sur $k$ redonne l'\'equi\-partition $B^2/\mu_0 \propto \rho V^2$. La contribution dominante \`a l'effet Joule provient de l'\'echelle $l_{\sigma}$. Nous obtenons
\begin{equation}
\frac{{\bf j}^2}{\sigma} = \frac{1}{\sigma} \int \vert \hat j \vert^2 \, dk \propto  \frac{1}{\mu_0^2 \sigma} \int k^2 \vert \hat B \vert^2 \, dk \propto  \frac{\rho}{\mu_0 \sigma} \,  \epsilon^{\frac{2}{3}} \,  l_{\sigma}^{-\frac{4}{3}} \propto  {\rho} \frac{V^3}{L},
\label{joulea}
\end{equation}
o\`u ${\bf j}$ est le vecteur densit\'e de courant.  
Nous constatons donc que la dissipation Joule est du m\^eme ordre que la puissance totale disponible. Remarquons qu'il en serait de m\^eme pour une dynamo de Batchelor suivant le sc\'enario (ii) pour $P_m \sim 1$, car bien que la densit\'e d'\'energie soit plus faible, l'\'echelle caract\'eristique du champ magn\'etique l'est \'egalement.

\selectlanguage{english}
% main text
\section{Introduction}
\label{Magnetic fields generated by turbulent flows}

It is now believed that magnetic fields of stars and possibly galaxies are generated by the motion of electrically conducting fluids through the dynamo process \cite{Zeldovitch,Widrow,Brandenburg}. These flows involve huge kinetic, $Re$, and magnetic, $R_m$, Reynolds numbers ($Re = VL/\nu$, $R_m = \mu_0 \sigma VL$, where $V$ is the $rms$ velocity amplitude, $L$ is the integral length scale, $\nu$ is the kinematic viscosity of the fluid, $\sigma$ is its electrical conductivity and $\mu_0$ is the magnetic permeability). No laboratory experiments, neither direct numerical simulations are possible in the range of $Re$ and $R_m$ involved in astrophysical flows. It is thus interesting to try to guess scaling laws for the magnetic field using some simple hypothesis. We consider here the minimum set of parameters, $V$, $L$, $\nu$, $\mu_0$, $\sigma$ and $\rho$, the fluid density. We note that discarding global rotation makes our results certainly invalid for many astrophysical objects but not all of them. Rotation is indeed not assumed important for the galaxies which do not display a large scale coherent magnetic field \cite{Zeldovitch,Widrow,Brandenburg}. Calling $B$ its $rms$ value, dimensional analysis gives
\begin{equation}
R_m^{c} = f(Re),
\label{threshold}
\end{equation}
for the dynamo threshold, and
\begin{equation}
\frac{B^2}{\mu_0} = \rho V^2 \, g(R_m, Re),
\label{saturation}
\end{equation}
for the mean magnetic energy density in the nonlinearly saturated regime. Our aim is to determine $f$ and $g$ in various regions of the parameter space $(R_m, Re)$, assuming that turbulence is homogeneous, isotropic and parity invariant (thus with no mean flow and no mean magnetic field generation through an alpha effect). As already mentioned, this may look like an academic exercise compared to most natural dynamos. It is however not more academic that the concept of homogeneous and isotropic turbulence with respect to real turbulent flows. We thus expect that our simple arguments may shed some light on open problems concerning the effect of turbulence on the dynamo threshold and on the dynamic equilibrium between magnetic and kinetic energy.

The dependence of the dynamo threshold $R_m^{c} = f(Re)$ in the limit of large $Re$ is still an open problem, even in the case of a homogeneous isotropic and parity invariant turbulent flow. Note that parity invariance prevents the generation of a large scale magnetic field via an alpha effect type mechanism and isotropy implies zero mean flow. Recent direct numerical simulations show that $R_m^{c}$ keeps increasing with $Re$ at the highest possible resolution without any indication of a possible saturation \cite{Schekochihin}.  Schekochihin et al. thus propose that two limit scenarios, sketched in figure 1, could be observed when computers will be able to reach higher $Re$: (i) saturation, $R_m^{c} \rightarrow \rm{constant}$, or (ii) increasing threshold in the form $R_m^{c} \propto Re$. 

\begin{figure}[!htb] 
\begin{center}
\includegraphics[width=.75\textwidth]{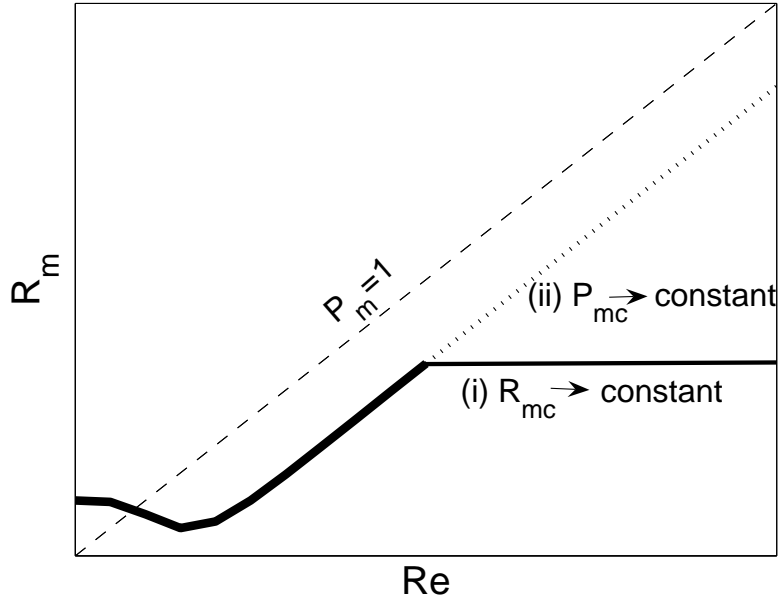}
\caption{Dependence of the dynamo threshold $R_{m}^c$  as a function of the Reynolds number $Re$. Scenario (i): $R_m^c$ tends to a constant. Scenario (ii):  $R_m^c$ is proportional to $Re$. }
\label{fig1}
\end{center}
\end{figure}

A lot of work has been performed on the determination of $R_m^{c}$ as a function of $Re$ for turbulent dynamos in the limit of large $Re$ (or small $P_m$). We recall that (ii) has been proposed  by Batchelor  in one of the first papers on turbulent dynamos \cite{Batchelor}. A lot of analytical studies have been also performed, mostly following Kazantsev's model \cite{Kazantsev} in order to show that purely turbulent flows can generate a magnetic field. Kazantsev considered a random homegeneous and isotropic velocity field, $\delta$-correlated in time and with a wave number spectrum of the form $k^{-p}$. He showed that for $p$ large enough, generation of a homogeneous isotropic magnetic field with zero mean value, takes place. This is a nice model but its validity is questionable for realistic turbulent flows. However, Kazantsev's model has been extrapolated to large $Re$. Various predictions, $R_m^{c} \propto Re$ \cite{Novikov}, $R_m^{c} \rightarrow \rm{constant} \approx 400$ for velocity spectra with $3/2 < p < 3$ and no dynamo otherwise \cite{Rogachevskii}, or dynamo for all possible slopes of the velocity spectrum $1 < p < 3$ \cite{Boldyrev} have been found. These discrepancies show that extrapolation of Kazantsev's model to realistic turbulence cannot be rigorous. The calculation is possible only in the case of a $\delta$-correlated velocity field in time, and $\delta (t-t')$, which has the dimension of the inverse of time, should then be replaced by a finite eddy turn-over time in order to describe large $Re$ effects. As already noticed, its choice is crucial to determine the behavior of $R_m^{c}$ versus $Re$.

A different problem about turbulent dynamos has been considered more recently. It concerns the effect of turbulent fluctuations on a dynamo generated by a mean flow. The problem is to estimate to which extent the dynamo threshold computed as if the mean flow were acting alone, is shifted by turbulent fluctuations. This question has been addressed only recently \cite{Fauve} and should not be confused with dynamo generated by random flows with zero mean. It has been shown that weak turbulent fluctuations do not shift the dynamo threshold of the mean flow at first order. In addition, in the case of small scale fluctuations, there is no shift at second order either, if the fluctuations have no helicity. This explains why the observed dynamo threshold in Karlsruhe and Riga experiments \cite{KarlsruheRiga} has been found in good agreement with the one computed as if the mean flow were acting alone, i.e. neglecting turbulent fluctuations. Recent direct numerical simulations have shown that in the presence of a prescribed mean flow, ${\overline{\bf v} (\bf { r})} \neq 0$, $R_m^{c}$ increases with $Re$ at moderate $Re$ but then seems to saturate at larger $Re$, thus following scenario (i). For the same flows, numerical modeling of large scales, large eddy simulations (LES) for instance, gives $R_m^{c} \sim \rm{constant}$  \cite{PontyLaval}. This last result follows from dimensional consideration as explained below, and has been also obtained for homogeneous isotropic turbulent non helical flows for which EDQNM closures have predicted $R_m^{c}  \approx 30$  \cite{Leorat}.

 \section{Turbulent dynamo threshold}
 \label{dynamo threshold}
 
 When the magnetic Prandtl number, $P_m = R_m / Re = \mu_0 \sigma \nu$, is small, $P_m \ll 1$,  the Ohmic dissipative scale, $l_{\sigma} = L R_m^{-3/4}$ is much larger than the Kolmogorov $l_K = L Re^{-3/4}$. Thus, if there is dynamo action, the magnetic field grows at scales much larger than $l_K $ and does not depend on kinematic viscosity. This hypothesis  is currently made for large scale quantities in turbulence and if correct, scenario (i) should be followed. If $\nu$ is discarded, $R_m^{c} = \rm{constant}$ indeed follows from dimensional analysis. It is thus not surprising that numerical models that do not resolve viscous scales, all gives this result, although the value of the constant seems to be strongly dependent on the flow geometry and on the model. 
We conclude that if dynamo action is observed for $P_m \ll 1$, the dynamo threshold is
\begin{equation}
R_m^{c} \rightarrow \rm{constant} \; \rm{when} \; Re \rightarrow \infty.
\label{thresholdsmallPm}
\end{equation}
However, we emphasize that no clear-cut demonstration of dynamo action by homogeneous isotropic and parity invariant turbulence exists for $P_m \ll 1$. Experimental demonstrations as well as direct numerical simulations all involve a mean flow and analytical methods extrapolated to $P_m \ll 1$ are questionable. 

It may be instructive at this stage to recall the study on turbulent dynamos made more than half a century ago by Batchelor \cite{Batchelor}. Using a questionable analogy between the induction and the vorticity equations, he claimed that the dynamo threshold corresponds to $P_m = 1$, i.e. $R_m^{c} \propto Re$, using our choice of dimensionless parameters (scenario (ii)). 
 
It is now often claimed that Batchelor's criterion $P_m > 1$ for the growth of magnetic energy in turbulent flows is incorrect. However, the weaker criterion $P_m > \rm{constant}$ (scenario (ii)) has not yet been invalidated by direct numerical simulations or by an experimental demonstration without mean flow.
 It is thus of interest to determine the minimal hypothesis for which Batchelor's predictions for dynamo onset is obtained using dimensional arguments. 
To wit, assume that the dynamo eigenmodes develop at small scales such that the threshold does not depend on the integral scale $L$. Then, discarding $L$ in our set of parameters, dimensional analysis gives at once $P_m = P_m^{c} = \rm{constant}$ for the dynamo threshold, i. e.
\begin{equation}
R_m^{c} \propto Re.
\label{thresholdbatchelor}
\end{equation}

It has been sometimes claimed that a non zero mean flow is necessary to get a dynamo following scenario (i). However, we note that even for a slow dynamo, i.e., growing on a diffusive time scale, the largest scales look stationary for a dynamo mode at wave length $l_{\sigma}$. For Kolmogorov turbulence, we indeed have, $\mu_0 \sigma l_{\sigma}^2/(L/V) \propto R_m^{-1/2} \ll 1$. This remains true for a $k^{-p}$ spectrum for $p<3$.

\section{Mean magnetic energy density}
\label{Mean magnetic energy density}
 
Dimensional arguments can be also used to determine scaling laws for the mean magnetic energy density. For $P_m \ll 1$ (scenario (i)), discarding $\nu$ gives 
\begin{equation}
\frac{B^2}{\mu_0} = \rho V^2 \, g_0(R_m),
\label{saturationsmallPm}
\end{equation}
where $g_0$ is an arbitrary function. Close to threshold, the $rms$ velocity $V$ is given by $\mu_0 \sigma VL \sim R_m^{c}$. In the case of a supercritical bifurcation, $g_0(R_m) \propto R_m - R_m^{c}$, and we obtain \cite{Petrelis01}
\begin{equation}
B^2 \propto \frac{\rho}{\mu_0 (\sigma L)^2} \, (R_m - R_m^{c}).
\label{saturationsmallPmthreshold}
\end{equation}
Far from threshold, $Re \gg R_m \gg R_m^{c}$, one could assume that $B$ no longer depends on $\sigma$ provided that the magnetic field mostly grows at scales larger than $l_{\sigma}$. We then obtain equipartition between magnetic and kinetic energy densities, 
\begin{equation}
B^2/\mu_0 \propto \rho V^2, 
\label{equipartition}
\end{equation}
as assumed by Biermann and Schl\"uter \cite{Biermann}.

A completely different result is obtained in scenario (ii). Let us first recall that according to Batchelor's analogy between magnetic field and vorticity \cite{Batchelor},  the magnetic field should be generated mostly at the Kolmogorov scale, $l_K = L Re^{-3/4}$, where the velocity gradients are the strongest. He then assumed that saturation of the magnetic field takes place for $\langle B^2 \rangle /\mu_0 \propto \rho v_K^2 = \rho V^2/\sqrt{Re}$, where $v_K$ is the velocity increment at the Kolmogorov scale, $v_K^2 = \sqrt{\nu \epsilon}$. $\epsilon = V^3/L$ is the power per unit mass, cascading from $L$ to $l_K$ in the Kolmogorov description of turbulence. 

This can be easily understood. $\epsilon = V^3/L$ being the power per unit mass available to feed the dynamo, it may be a wise choice to keep it, instead of $V$ in our set of parameters, thus becoming $B$, $\rho$, $\epsilon$, $L$, $\nu$, $\mu_0$ and $\sigma$. Then, if we consider dynamo modes that do not depend on $L$, we obtain at once 
\begin{equation}
\frac{B^2}{\mu_0} = \rho \sqrt{\nu \epsilon} \, h (P_m) = \frac{\rho V^2}{\sqrt{Re}} \, h (P_m)
\label{saturationbatchelor}
\end{equation}
for saturation, where $h(P_m)$ is an arbitrary function of $P_m$. Close to dynamo threshold, $P_m \approx P_m^{c}$, we have $h(P_m) \propto P_m - P_m^{c}$ if the bifurcation is supercritical. Only the prefactor $\rho V^2 / \sqrt{Re}$ of (\ref{saturationbatchelor}) is the kinetic energy at Kolmogorov scale, that was assumed to be in equipartition with magnetic energy in Batchelor's prediction. This class of dynamos being small scale ones, it is not surprising that the inertial range of turbulence screens the magnetic field from the influence of integral size, thus $L$ can be forgotten. We emphasize that a necessary condition for Batchelor's scenario is that the magnetic field can grow below the Kolmogorov scale, i.e. its dissipative length $l_{\sigma}$ should be smaller than $l_K$, thus $P_m > 1$. 

There is obviously a strong discrepancy between (\ref{equipartition}) and (\ref{saturationbatchelor}). 
The prefactors in these two laws are the upper and lower limits of a continuous family of scalings that are obtained by balancing the magnetic energy with the kinetic energy at one particular length scale within the Kolmogorov spectrum. It is not known if one of them is selected by turbulent dynamos. 

 \section{Ohmic losses}
 
Ohmic losses due to currents generated by dynamo action give a lower bound to the power required to feed a dynamo. In order to evaluate them, it is crucial to know at which scales the magnetic field grows.
Assuming that a dynamo is generated in the case $P_m \ll 1$ (scenario (i)), we want to give a possible guess for the power spectrum $\vert \hat B \vert^2$ of the magnetic field as a function of the wave number $k$ and the parameters $\rho$, $\epsilon$, $L$, $\nu$, $\mu_0$ and $\sigma$. Far from threshold,  $Re \gg R_m \gg R_m^{c}$, the dissipative lengths are such that $l_K \ll l_{\sigma} \ll L$. For $k$ in the inertial range, i.e. $k l_{\sigma} \ll 1 \ll kL$, we may use a Kolmogorov type argument and discard $L$, $\sigma$ and $\nu$. Then, only one dimensionless parameter is left, and not too surprisingly, we get
\begin{equation}
\vert \hat B \vert^2 \propto \mu_0 \rho \,  \epsilon^{\frac{2}{3}} \, k^{-\frac{5}{3}}.
\label{kolmogorov}
\end{equation}
This is only one possibility among many others proposed for MHD turbulent spectra within the inertial range, but it is the simplest. Integrating over $k$ obviously gives the equipartition law  (\ref{equipartition}) for the magnetic energy. It is now interesting to evaluate Ohmic dissipation. Its dominant part comes from the current density at scale $l_{\sigma}$. We have
\begin{equation}
\frac{{\bf j}^2}{\sigma} = \frac{1}{\sigma} \int \vert \hat j \vert^2 \, dk \propto  \frac{1}{\mu_0^2 \sigma} \int k^2 \vert \hat B \vert^2 \, dk \propto  \frac{\rho}{\mu_0 \sigma} \,  \epsilon^{\frac{2}{3}} \,  l_{\sigma}^{-\frac{4}{3}} \propto  {\rho} \frac{V^3}{L}.
\label{joule}
\end{equation}
We thus find that Ohmic dissipation is proportional to the total available power which corresponds to some kind of optimum scaling law for Ohmic dissipation. Although, this does not give any indication that this regime is achieved, we note that the above scaling corresponds to the one found empirically from a set of numerical models \cite{christensen}. Their approximate fit, $(B^2/\mu_0)/(j^2/\sigma) \propto L/V$, indeed results from equations (\ref{kolmogorov}, \ref{joule}).

% \appendix

% \section{}
% \label{}

% The Acknowledgements are also a un-numbered section
%\section*{Acknowledgements}
% Acknowledgements text here

\end{document}